\documentclass[%
 aip, amsmath,amssymb, reprint, floatfix%
]{revtex4-1}
\usepackage{graphicx, dcolumn, bm, mathptmx, etoolbox, xcolor, lipsum, siunitx, soul, makecell}
\usepackage[utf8]{inputenc}
\usepackage[T1]{fontenc}
\usepackage[normalem]{ulem}
\DeclareUnicodeCharacter{221A}{$\sqrt{}$}
\DeclareUnicodeCharacter{2202}{$\partial$}

\DeclareSIUnit\oersted{Oe}
\newcommand{\ipc}{$\mathrm{ions/cm^2}$}

\makeatletter
\def\@email#1#2{%
 \endgroup
 \patchcmd{\titleblock@produce}
  {\frontmatter@RRAPformat}
  {\frontmatter@RRAPformat{\produce@RRAP{*#1\href{mailto:#2}{#2}}}\frontmatter@RRAPformat}
  {}{}
}%
\makeatother
\begin{document}

\preprint{AIP/123-QED}

\title{Structural Properties and Recrystallization Effects in Ion Beam Modified B20-type FeGe Films}
%
%

\author{Jiangteng Liu}
\affiliation{Department of Electrical \& Computer Engineering, University of Washington, Seattle, WA 98195}%

\author{Ryan Schoell}%
\affiliation{Sandia National Laboratories, Albuquerque, NM 87123}%

\author{Xiyue S. Zhang}%
\affiliation{Department of Applied Physics, Cornell University, Ithaca, NY 14853}%

\author{Hongbin Yang}%
\affiliation{Department of Applied Physics, Cornell University, Ithaca, NY 14853}%

\author{M. B. Venuti}
\affiliation{Department of Physics, Colorado School of Mines, Golden, CO 80401}%
 
\author{Hanjong Paik}
\affiliation{Platform for the Accelerated Realization, Analysis, and Discovery of Interface Materials, Cornell University, Ithaca, NY 14853}
\affiliation{School of Electrical \& Computer Engineering, University of Oklahoma, Norman, OK 73019}
\affiliation{Center for Quantum Research and Technology, University of Oklahoma, Norman, OK 73019}

\author{David A. Muller}%
\affiliation{Department of Applied Physics, Cornell University, Ithaca, NY 14853}%

\author{Tzu-Ming Lu}%
\affiliation{Center for Integrated Nanotechnologies, Sandia National Laboratories, Albuquerque, NM 87123}%

\author{Khalid Hattar}%
\affiliation{Department of Nuclear Engineering, University of Tennessee, Knoxville, TN 37996} 

\author{Serena Eley}
\affiliation{Department of Electrical \& Computer Engineering, University of Washington, Seattle, WA 98195}
\affiliation{Department of Physics, Colorado School of Mines, Golden, CO 80401}

\email[Corresponding author: ]
{serename@uw.edu}

\date{\today}

\begin{abstract}
Disordered iron germanium (FeGe) has recently garnered interest as a testbed for a variety of magnetic phenomena as well as for use in magnetic memory and logic applications. This is partially owing to its ability to host skyrmions and antiskyrmions --- nanoscale whirlpools of magnetic moments that could serve as information carriers in spintronic devices. In particular, a tunable skyrmion-antiskyrmion system may be created through precise control of the defect landscape in B20-phase FeGe, motivating developing methods to systematically tune disorder in this material and understand the ensuing structural properties.  To this end, we investigate a route for modifying magnetic properties in FeGe. Specifically, we irradiate epitaxial B20-phase FeGe films with 2.8 MeV Au$^{4+}$ ions, which creates a dispersion of amorphized regions that may preferentially host antiskyrmions at densities controlled by the irradiation fluence. To further tune the disorder landscape, we conduct a systematic electron diffraction study with in-situ annealing, demonstrating the ability to recrystallize controllable fractions of the material at temperatures ranging from approximately 150$^{\circ}$ C to 250$^{\circ}$C.  Finally, we describe the crystallization kinetics using the Johnson-Mehl-Avrami-Kolmogorov model, finding that the growth of crystalline grains is consistent with diffusion-controlled one-to-two dimensional growth with a decreasing nucleation rate.

\end{abstract}

\maketitle

\section{Introduction}\label{sec:intro}

Iron germanium (Fe-Ge) is an excellent candidate for use in thermoelectric generators, owing to its high thermoelectric figure of merit,\cite{Sato2016, Verchenko2017} and spintronics, due to its magnetic properties. Depending on its crystalline structure, FeGe can exhibit collinear or canted antiferromagnetic order,\cite{OBeckman1972} a coexisting charge-density wave and antiferromagnetic phase,\cite{Teng2022} as well as helical, conical, and skyrmion lattice phases.\cite{Yu_2010} Under atmospheric pressure, stoichiometric FeGe crystallizes into three different polymorphs: cubic structure (B20-phase) with P2$_1$3 symmetry, hexagonal kagome structure (B35-phase) with P6/mmm symmetry, and a monoclinic lattice. The cubic B20-phase stabilizes at growth temperatures below 580 $^{\circ}$C, above which the hexagonal and monoclinic phases may form.  Off-stoichiometric variants may also appear in FeGe crystals and films.  On the Ge-rich side of the temperature-composition phase diagram, shown in Fig. \ref{fig:fig1}(a), Fe$_2$Ge$_3$ can be stabilized at temperatures below 600 $^{\circ}$C, whereas several binary compounds may form on the Fe-rich side of the phase diagram, including Fe$_3$Ge and Fe$_2$Ge.\cite{VillarsP, Fe3Ge, D1CE00970B}

B20-phase cubic FeGe has recently attracted considerable interest due to the discovery of skyrmions in this polymorph. Skyrmions are topologically protected chiral spin textures that are being evaluated as potential information carriers in low-energy spintronic devices. These textures originate from an antisymmetric exchange interaction called the Dzyaloshinskii-Moriya interaction (DMI) that can arise in magnetic materials with strong spin-orbit coupling and broken inversion symmetry. Unlike the exchange interaction that leads to ferromagnetic order (a collinear spin arrangement), the DMI can tilt magnetic moments such that the combined interactions engender whirlpool-like spin structures that behave as particles that can be controlled by applied currents.\cite{Yu_2010, Nagaosa2013, Yu2012, Fert2013, Jonietz2010}

Skyrmions were first observed in B20-phase cubic materials, such as MnSi,\cite{Muhlbauer2009} Fe$_{1-x}$Co$_x$Si,\cite{Yu2010LTEM} and FeGe,\cite{Yu_2010} due to isotropic DMI originating from the non-centrosymmetric crystal structure.  On the other hand, if the DMI is anisotropic, non-symmetric skyrmions or antiskyrmions --- the antiparticles of skyrmions --- can form.\cite{Hoffmann2017} Skyrmion-antiskyrmion systems are of interest for binary data encoding in racetrack memory and logic applications as well as testbeds for studying the dynamics of skyrmion-antiskyrmion crystals and liquids\cite{PhysRevB.93.064430, Leonov2015} as well as skyrmion-antiskyrmion pair annihilation,\cite{Hu2017} which is predicted to produce a spin wave. \cite{Zhang2017b}  Few materials have been identified in which these particles coexist,\cite{PhysRevB.106.014421} namely in Mn$_2$RhSn,\cite{sivakumar} Co/Ni multilayers,\cite{Hassan2024}  a 70-nm-thick FeGe single crystal,\cite{Zheng2022} and chemically disordered amorphous 80 nm-thick Fe$_x$Ge$_{1-x}$ films.\cite{Streubel} In fact, in a recent study,\cite{Eley2024} we provided evidence for the coexistence of skyrmions and antiskyrmions in epitaxial B20-phase FeGe films by ion beam modification, which induced amorphous regions within the crystalline matrix. Low-temperature electrical transport and magnetization studies revealed a strong topological Hall effect with a double-peak feature that is consistent with recognized signatures of skyrmions and antiskyrmions, especially in light of their observation in disordered FeGe through Lorentz transmission electron microscopy (LTEM) and x-ray magnetic circular dichroism spectroscopy (XMCD).\cite{Streubel}

In this study, we introduce another potential avenue for tunability of spin structures in FeGe. We first ion-beam modify FeGe at different fluences. Second, through in-situ annealing and selective area electron diffraction, we demonstrate the ability to systematically recrystallize the amorphized regions to the target B20-phase and study the associated recrystallization kinetics.  By analyzing the recrystallization kinetics using the Johnson–Mehl–Avrami–Kolmogorov (JMAK) model, we describe the dynamic processes that govern the formation of crystalline regions from the amorphous matrix. This knowledge not only enhances our ability to fine-tune the defect landscape for skyrmion and antiskyrmion stabilization but also provides insights into the thermal stability and transformation pathways of these and other topological structures that form in disordered FeGe.

\section{Results and Discussion}\label{sec:results}

\subsection{\label{sec:level2}Growth and Ion-Beam Modification of FeGe Films}

\begin{figure}[h]
\centering
\includegraphics[width=1\linewidth]{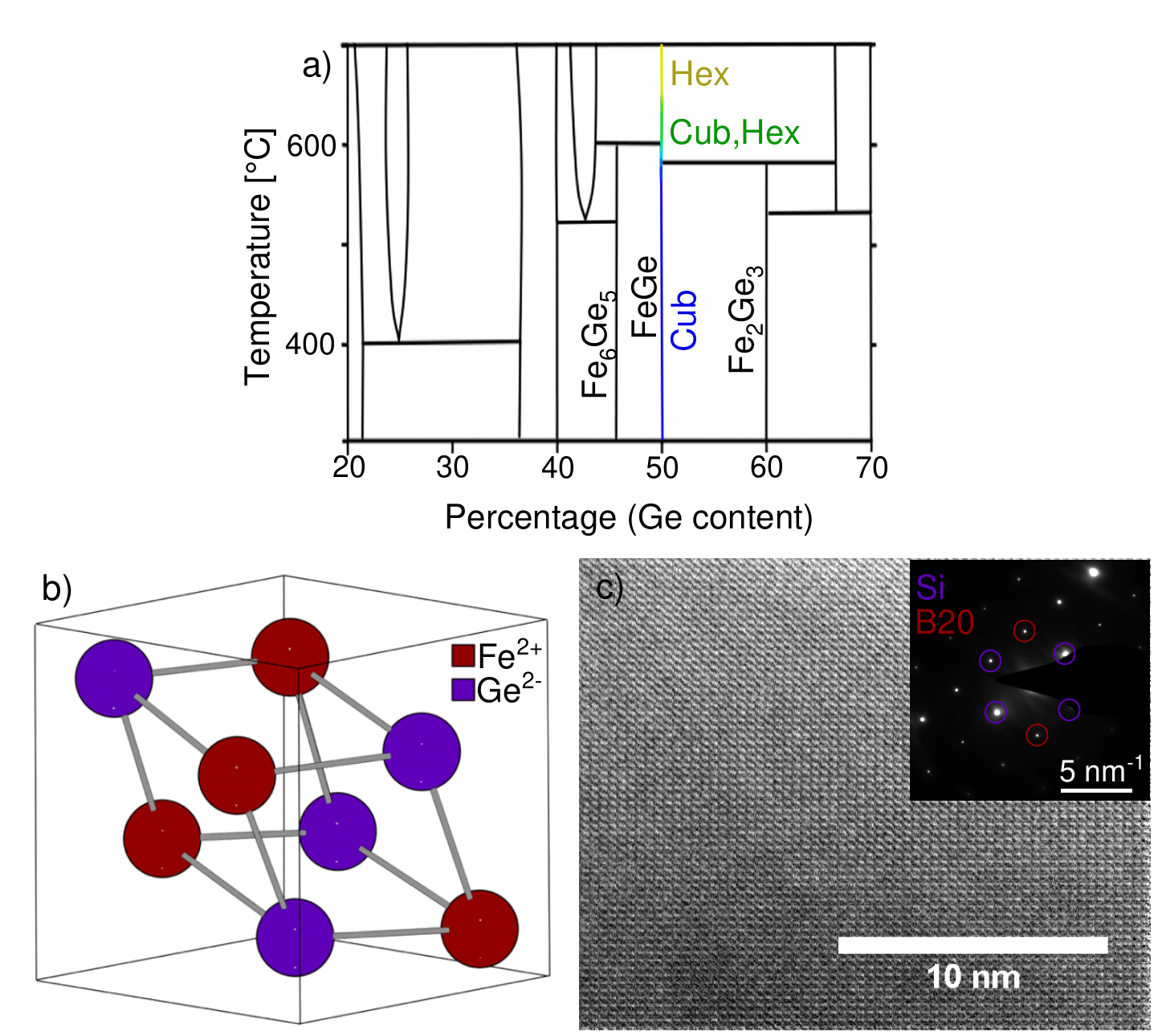}
\caption{\label{fig:fig1}  (a) Growth temperature - composition phase diagram of the Fe-Ge system, adapted from Refs.~[\onlinecite{FeGephase1, FeGephase2}]. (b) Diagram of B20-phase crystal structure for FeGe. (c) High-angle annular dark field STEM image of the as-grown FeGe film layer with a zone axis of [211]. Inset shows the SAED pattern of the as-grown FeGe; bright spots associated with Si and the B20-phase of FeGe are identified.}
\end{figure}

Using molecular beam epitaxy, we grew approximately 55-nm thick epitaxial films of FeGe in the B20-phase, which has space group \textit{P}2$_1$3 with eight atoms per unit cell and cubic structure, as shown in Fig.~\ref{fig:fig1}(b). To characterize the material structure, we performed x-ray diffraction (XRD), scanning transmission electron microscopy (S/TEM), and selected area electron diffraction (SAED) on the films or lamellae prepared by focused ion beam microscopy.  Details regarding the materials growth procedure, STEM characterization, and lamella preparation are included in the Methods section. XRD spectra (shown in Ref.~[\onlinecite{Eley2024}]) reveal a peak at $2\theta =33.1 ^{\circ}$ that is consistent with the FeGe (111) B20-phase. Figure~\ref{fig:fig1}(c) displays an annular dark field cross-sectional STEM image of the as-grown FeGe film and the inset shows the corresponding SAED pattern. In general, SAED images show a two-dimensional (2D) slice of the reciprocal lattice, resulting in sharp spots from reflections off lattice planes.  Single crystals with little disorder produce only sharp spots, whereas defined and diffuse rings result from nanocrystalline powders and amorphous samples, respectively. Here, we see only bright diffraction spots, revealing a highly ordered crystalline structure, with no observable diffuse ring. 

\begin{figure}[htb!]
\centering
\includegraphics[width=1\linewidth]{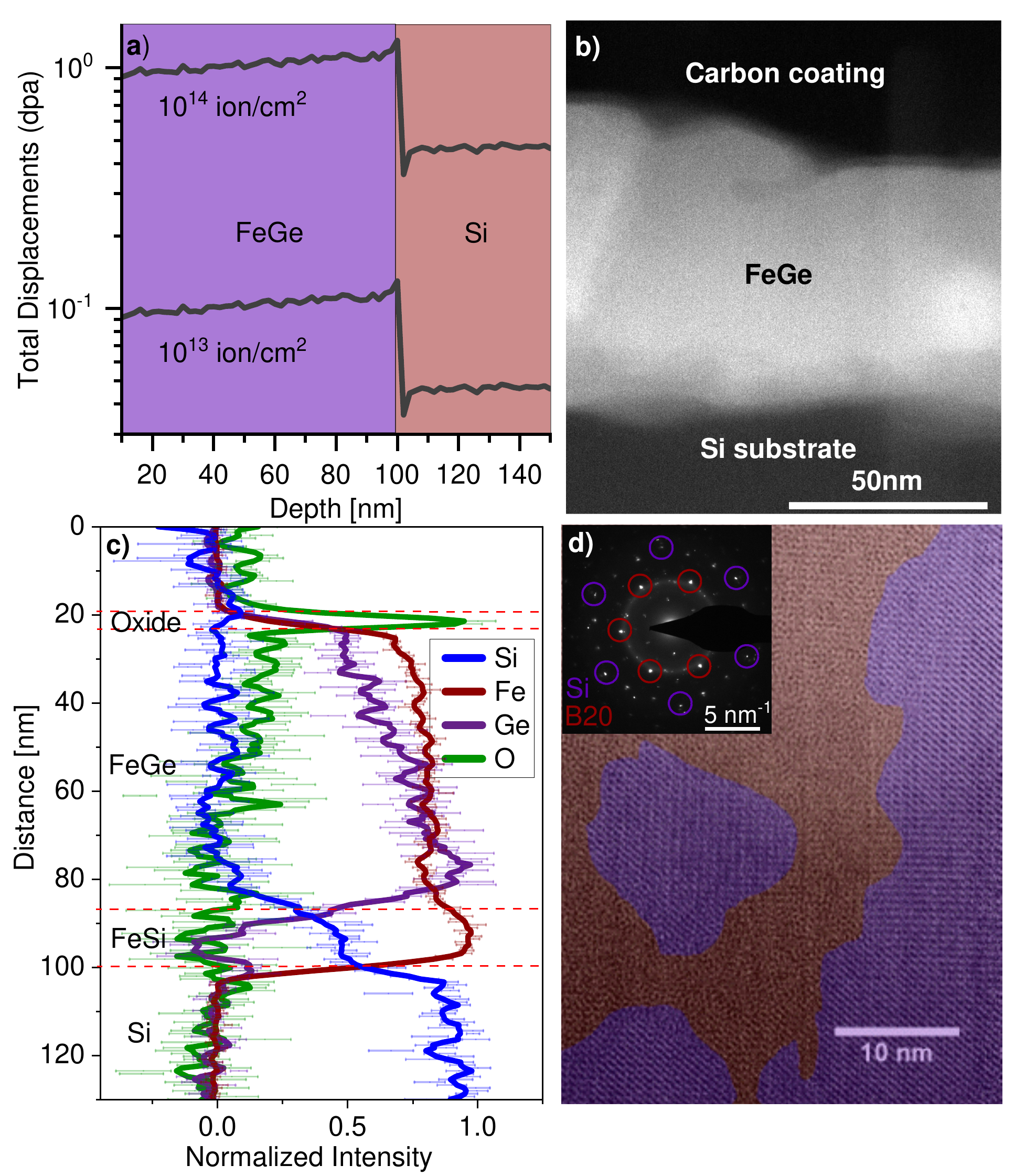}
\caption{\label{fig:fig2} (a) SRIM simulations displacement profiles for FeGe irradiated at $10^{13}$ and $10^{14}$ \ipc. (b) High angle annular dark field (HAADF) cross-sectional STEM image, (c) EELS profile, and (d) high-resolution annular bright-field STEM image of a cross-section of the FeGe film irradiated at $10^{13}$ \ipc. (d)  False color overlay highlighting the amorphous (red) and crystalline (purple) regions. Inset is the SAED pattern for the same irradiated sample before annealing; bright spots associated with Si and the FeGe B20-phase are circled.}
\end{figure}

We subsequently irradiated cleaved \SI{6}{\milli\meter} $\times$ \SI{6}{\milli\meter} cuts of the films with 2.8 MeV Au$^{4+}$ ions using a 6 MV High Voltage Engineering (HVE) EN Tandem Van de Graaff Accelerator at the Sandia National Laboratories Ion Beam Lab.  To test the effects of various defect levels, each film was irradiated with a fluence of 10$^{13}$ and 10$^{14}$ \ipc, inducing 10$^{-1}$ and 1 displacement per atom (dpa), respectively, as determined by the Stopping and Range of Ions in Matter (SRIM) simulations, shown in  Fig.~\ref{fig:fig2}(a). The SRIM results predict that the irradiation process should introduce a homogeneous distribution of defects throughout the film depth, without implanting Au-ions into the FeGe. Further details of the ion beam modification process and corresponding SRIM simulations are included in the Methods section.

To identify phases as well as determine the effects of irradiation on the crystalline structure, chemical composition throughout the film depth, and crystalline orientation, we performed STEM imaging, electron energy loss spectroscopy (EELS), and SAED. The low magnification cross-sectional STEM image in Fig.~\ref{fig:fig2}(b) shows a top-layer carbon coating (applied for microscopy), FeGe layer, and Si (111) substrate. In addition to these three layers, the EELS results in Fig.~\ref{fig:fig2}(c) show a surface oxide, the thin FeSi seed layer that was formed on the substrate to mediate growth of epitaxial B20-phase FeGe, no detectable Au concentration within the measurement resolution (consistent with simulations that indicate Au was not implanted in the FeGe). In the high-resolution annular bright-field STEM image in Fig. \ref{fig:fig2}(d) of a film irradiated at $10^{13}$ \ipc \, we also see irradiation-induced amorphous regions within the crystalline matrix, highlighted using red and purple false color overlays, respectively. The presence of amorphous regions is also revealed in the SAED pattern, in which a distinct diffuse ring is notable that did not exist in the SAED image of the pristine sample [Fig.~\ref{fig:fig1}(c) inset].

\subsection{\label{sec:recrystalize}In-situ recrystallization of Irradiated FeGe}

\begin{figure}[ht!]
\includegraphics[width=0.7\linewidth]{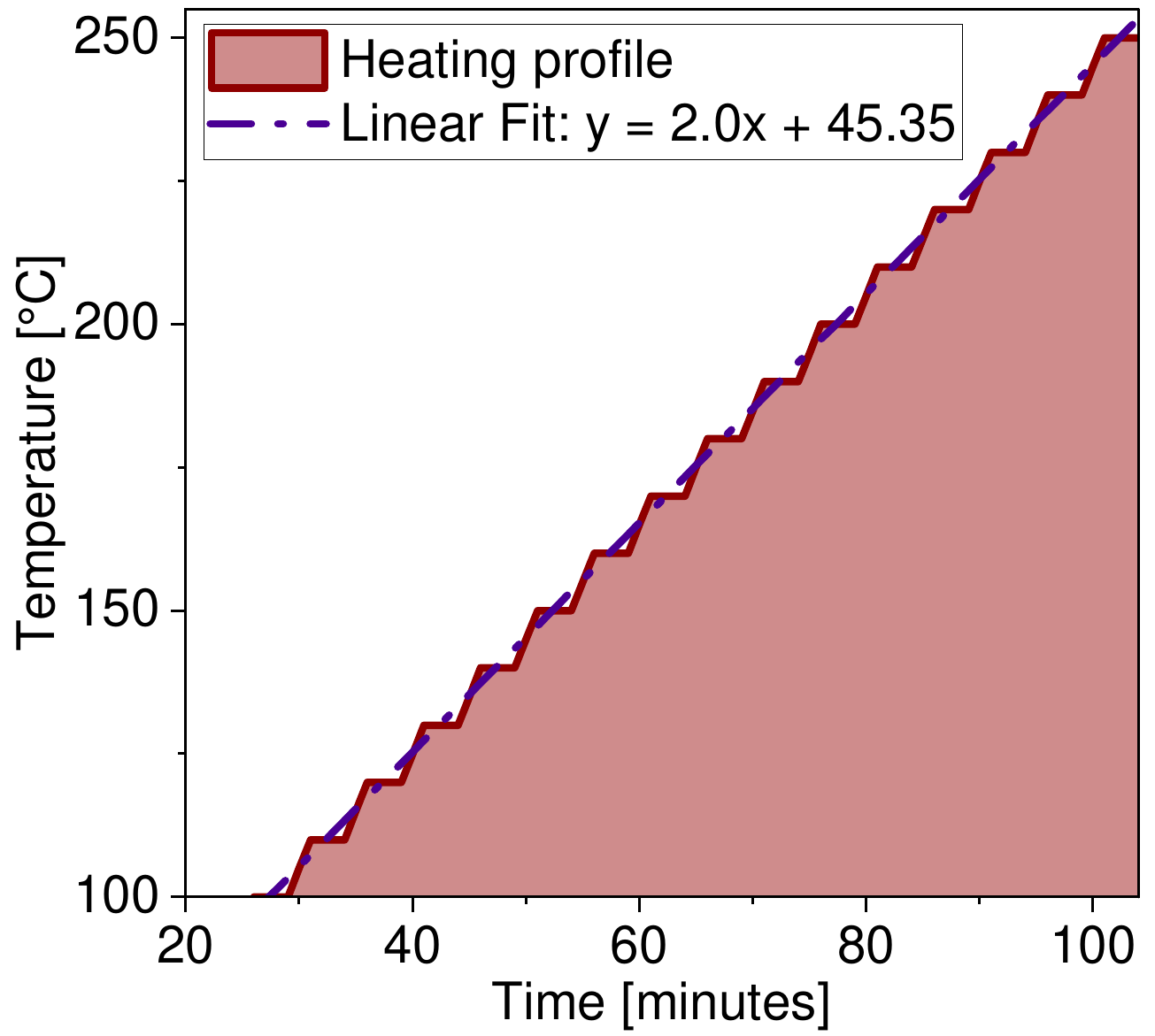}
\caption{\label{fig:fig3} Heating profile used in the select area electron diffraction study with in-situ annealing. The red line shows the step-wise heating curve. The dashed dark purple line is the linear approximation of the heating profile used when applying the extended (non-isothermal) Johnson–Mehl–Avrami–Kolmogorov (JMAK) model to understand the recrystallization process.}
\end{figure}

The next goal was to develop an annealing procedure to tune the density and sizes of the amorphized versus crystalline regions in the irradiated FeGe films, which would subsequently enable tunability of the relative skyrmion and antiskyrmion populations. To actively monitor the crystalline orientation, we performed SAED with in-situ annealing on plan view lamellae of films irradiated at $10^{13}$ \ipc \ and $10^{14}$ \ipc, as well as a pristine lamella, for reference.  The electron beam was incident on the FeGe (111) lattice plane and the annealing process followed the stepwise heating profile depicted in Fig. \ref{fig:fig3}. Further details describing the experimental setup are included in the Methods section and SAED data for the pristine film is presented in the Supplementary Materials.

\begin{figure*}[p]
\centering
\includegraphics[width=0.67\linewidth]{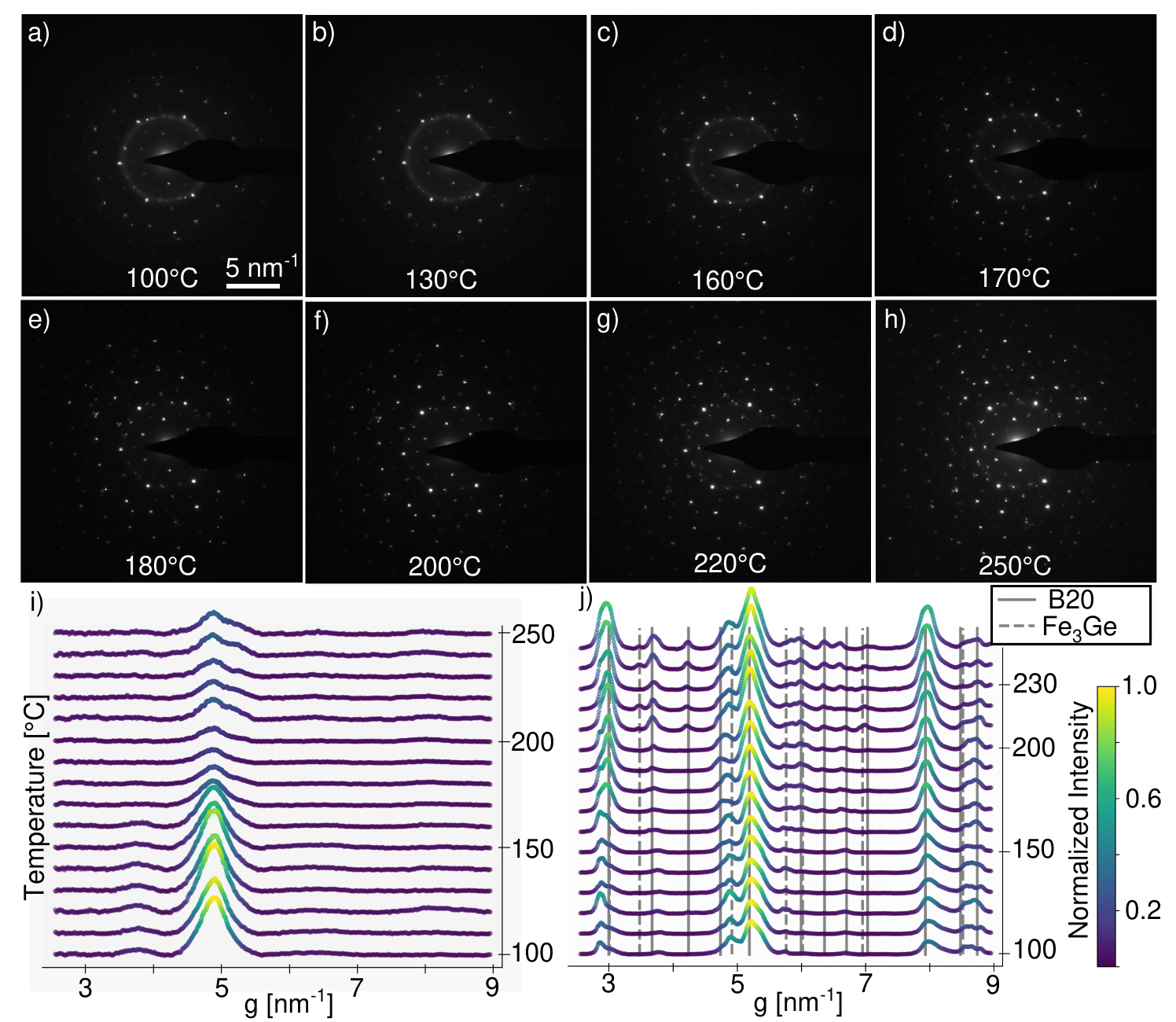}
\caption{\label{fig:fig4} Selected area electron diffraction images of FeGe film irradiated at $10^{13}$ \ipc. (a-h) Raw electron diffraction images at different stages of annealing, at the indicated temperatures. Extracted intensity (color scale) versus reciprocal space parameter $g$ for the (i) amorphous regions and (j) crystalline regions at different annealing temperatures. Curves offset for clarity. The solid and dashed grey lines label the expected position of peaks associated with the B20-phase of FeGe and cubic Fe$_3$Ge phase, respectively.}

\centering
\includegraphics[width=0.67\linewidth]{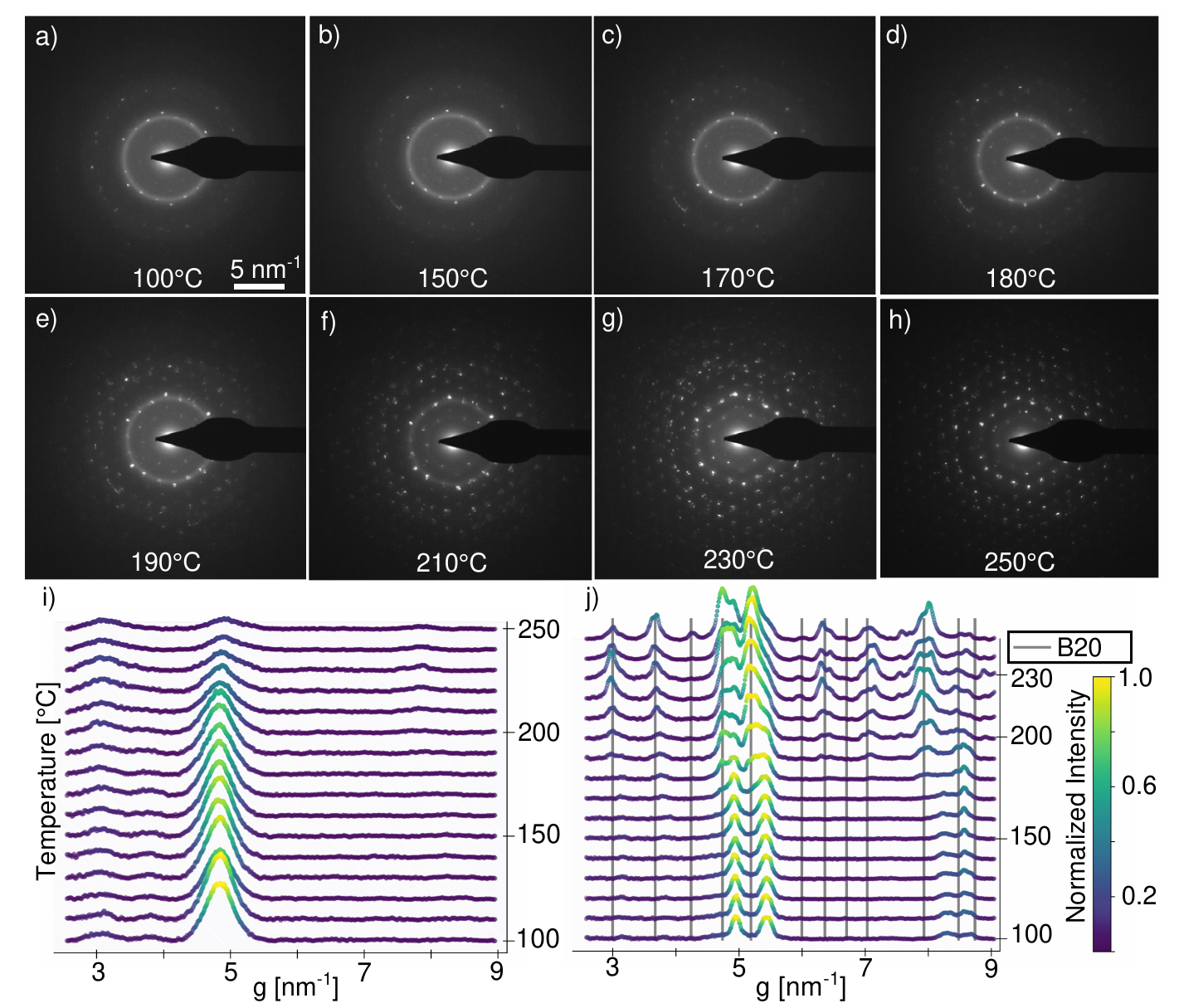}
\caption{\label{fig:fig5} Selected area electron diffraction images of FeGe film irradiated at $10^{14}$ \ipc. (a-h) Raw electron diffraction images at different stages of annealing, at the indicated temperatures. Extracted intensity (color scale) versus reciprocal space parameter $g$ for the (i) amorphous regions and (j) crystalline regions at different annealing temperatures. Curves offset for clarity.  The solid grey lines indicate the expected position of peaks associated with the FeGe B20-phase phase.}
\end{figure*}

Figure \ref{fig:fig4}(a-h) presents the raw SAED patterns of the sample irradiated at $10^{13}$ \ipc \, with each panel showing results at a fixed location while heating at the indicated temperature. We see that as the temperature rises, the number of diffraction spots noticeably increases while the intensity of the diffuse ring diminishes, a trend suggestive of heating-induced recrystallization of the amorphous regions. To approximate the relative fraction of crystalline regions and the effect of each heating stage on this fraction, we separate the raw data into amorphous and crystalline components, and then integrate to convert the 2D diffraction data into one-dimensional (1D) profiles of normalized intensity. Finally, the intensity data is plotted against the reciprocal space parameter $g = 1/d$, where $d$ is the spacing between crystalline reflection planes. Refer to the Supplementary Material for more details on our signal separation procedure.

Figures~\ref{fig:fig4}(i) and \ref{fig:fig4}(j) display the resulting separated 1D profiles for the amorphous and crystalline components, respectively. For the amorphous component [Fig.~\ref{fig:fig4}(i)], the diffuse ring at $1/d = 4.9\ \mathrm{nm}^{-1}$ matches expected values for amorphous FeGe, as reported in previous studies,\cite{AmorphousFeGe} and the peak intensity decreases with increasing temperature. On the other hand, Fig. \ref{fig:fig4}(j) illustrates an increase in the number and intensity of diffraction peaks associated with the crystalline component as the temperature rises. To identify phases associated with these peaks, we used the Materials Project database\cite{MaterialsProject_FeGe} of calculated and experimental diffraction pattern data for the Fe-Ge and Fe-Ge-O systems, and subsequently found that these peaks correspond to the B20-phase of FeGe as well as to the cubic Fe$_{3}$Ge-$Fm\overline{3}m$ phase. More information regarding the phase identification process is discussed in the Supplementary Materials. 

We repeated the in-situ annealing and SAED process on the sample irradiated at the higher fluence of $10^{14}$ \ipc, observing similar recrystallization behavior, as shown in Fig.~\ref{fig:fig5}. It is important to note that the peak intensities in the SAED patterns are influenced by the image contrast during data acquisition. Though this was not adjusted during the annealing process, it did differ between runs for each sample such that peak intensities should not be compared between the different samples. As expected, the width and relative intensity of the diffuse ring (compared to the diffraction spots) are significantly stronger than observed in the sample irradiated at a lower fluence. This can be attributed to the higher irradiation dose inducing larger amorphous regions. Additionally, the greater number of diffraction spots suggests that the recrystallized grains possess more grain boundaries with varying orientations.

We again separate the amorphous and crystalline components into 1D profiles of intensity versus $g$, shown in Figs.~\ref{fig:fig5}(i) and \ref{fig:fig5}(j), respectively. Similar to the other irradiated film, Fig.~\ref{fig:fig5}(i) reveals that the intensity of the amorphous component decreases with temperature, but we also see that the diffuse ring persists over a wider range of temperatures. This indicates that the initial higher density of amorphous regions in the sample irradiated at $10^{14}$ \ \ipc \ results in increased stability of the amorphous phase during annealing.
Regarding the crystalline component, shown in Fig. \ref{fig:fig5}(j), we again see an increase in the number and intensity of diffraction peaks with temperature. In this case, the identifiable phase is primarily B20 FeGe. However, additional unidentified phases are observed at temperatures between 100-180$^{\circ}$C at a $1/d$ of 4.9 $\mathrm{nm}^{-1}$ and 5.5 $\mathrm{nm}^{-1}$, which may originate from lattice strain, oxides, or various other defects such as twin boundaries and stacking faults.\cite{Latticeshift, twinboundries, Superlattice, DiffractionArtifcate} 

\subsection{\label{sec:kinetics}Recrystallization Kinetics of Irradiated FeGe}

To determine the evolution of the effective crystalline volume fraction with temperature and time, we characterize the recrystallization kinetics using the Johnson–Mehl–Avrami–Kolmogorov (JMAK) model.\cite{Johnson1939ReactionKI} This model provides an estimate of the local Avrami exponent $n$, which can be used to extract information about nucleation and grain growth\cite{AvramiConstant} --- the nucleation rate at different stages of crystallization (increase or decrease), the growth mechanism (interface or diffusion controlled), and the dimensionality of the grain growth in the system (3D-bulk materials, 2D-thin film, 1D-nanotube). The classic JMAK model assumes that the phase transformation occurs under isothermal conditions, low anisotropy during crystal growth, and a random distribution of nucleation sites in the parent phases.\cite{JMAKbackground} The model finds that the time $t$ evolution of the effective crystalline volume fraction $X$ can be expressed as\cite{Johnson1939ReactionKI}

\begin{eqnarray}\label{eq:JMAKclassic}
    X(t) = 1 - \text{exp}[-k(t-t_{0})^{n}],
\end{eqnarray} where $k$ is the kinetic coefficient, which is constant under a fixed temperature, and $t_0$ is induction time (the duration between when sample heating initiates and the onset of crystallization).\cite{Blazquez2022} Consequently, the local Avrami exponent depends on the crystalline volume fraction as follows: \cite{NonIsoApplic, NonIsoModel}

\begin{eqnarray}\label{eq:JMAKclassicAvrami}
     n(X) &=& \frac{\mathrm{d}\ln[-\ln(1-X)]}{\mathrm{d}\ln(t-t_0)}~.
\end{eqnarray}

\noindent In our system, all the phases involved are cubic, which is highly symmetric. Also, we assume that the ion-beam modification process generates an even distribution of amorphous regions within the FeGe sample, which supports the model's applicability in our study. However, because our stepwise heating profile alternates between isothermal steps and temperature ramps, we also consider the extended JMAK model applicable for non-isothermal conditions. In this model, the crystalline volume fraction depends on the Avrami exponent, time, and temperatures as\cite{Nakamura1, Nakamura2}

\begin{eqnarray}\label{eq:JMAKextended_vs_time}
    X_{extended}(t) = 1 - \exp\left\{ -\left[ \int_{t_{0}}^{t} k(T) \, dt \right]^n \right\}.
\end{eqnarray}

\noindent For a constant heating rate where $dT/dt \equiv \beta$, the term $\int^{t}_{t_{0}} k(T) \ dt$ can be rewritten as $\int^{T}_{T_{0}} k(T)/\beta \ dT$ such that  $\int^{T}_{T_{0}} k(T)dT = k'(T-T_{0})$.\cite{NonIsoModel, NonIsoApplic} Hence, the temperature dependence of the crystalline volume fraction becomes

\begin{eqnarray}\label{eq:JMAKextended_vs_temp}
    X_{extended}(T) &=& 1 - \exp\left\{- \left[\frac{k'(T-T_{0})}{\beta}\right]^n \right\} 
\end{eqnarray}

\noindent for $k'(T) = k'_{0}\exp(\frac{E_{a}}{RT})$, $k'_{0}$ is a constant, $E_a$ is activation energy, $T_0$ is the onset temperature for crystallization, and $R$ is the gas constant. Rearranging terms and taking the logarithm of both sides of the equation, we see that the local Avrmai constant under non-isothermal conditions can be expressed as\cite{NonIsoModel, NonIsoApplic}

\begin{eqnarray}\label{eq:JMAKextendedAvrami}
    n_{extended}(X) &=& \left[\frac{RT^{2}}{RT^{2}+E_a(T-T_0)} \right] \frac{\mathrm{d}\ln[-\text{ln}(1-X)]}{\mathrm{d}\ln[(T-T_0)/\beta]}.
\end{eqnarray}

Based on Eqs.~(\ref{eq:JMAKclassicAvrami}) and (\ref{eq:JMAKextendedAvrami}), we calculated the local Avrami exponent using both the classic and extended JMAK model to describe the nucleation and growth mechanism of our irradiated FeGe samples. To apply these models, the effective crystalline volume fraction $X$ must first be obtained. Having separated the diffraction data into amorphous and crystalline components, we can apply a classic Rietveld-based metric called the Degree of Crystallization (DOC) to extract the crystalline volume fraction.\cite{DegreeOfCrystalline, MadsenScarlettKern2011, Fauziyah_2022} Here, DOC is simply defined as the area-under-the-curve of the diffraction intensity for the crystalline component $I_{cryst}(g)$ normalized by the area-under-the-curve for the total diffraction intensity --- the sum of $I_{cryst}(g)$ and the intensity of the amorphous component, $I_{amorph}(g)$: 

\begin{eqnarray}\label{eq:DOC}
\text{DOC} = \frac{\int{I_{cryst}}dg}{\int{I_{cryst}}dg + \int{I_{amorph}}dg} 
\end{eqnarray}

\begin{figure}[h!]
\centering
\includegraphics[width=1\linewidth]{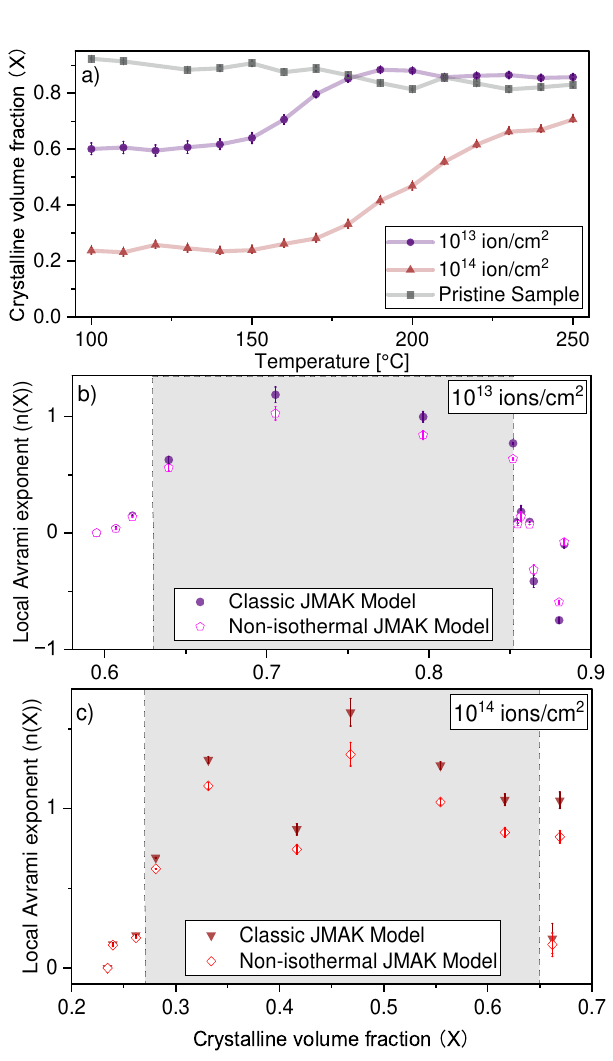}
\caption{\label{fig:fig6} (a) Temperature-dependent effective crystalline volume fraction $X$ for the irradiated FeGe films, where $X$ is calculated using [Eq. (\ref{eq:DOC})] from the processed diffraction data. (b-c) Local Avrami exponent plotted against effective crystalline volume fraction obtained using both the classical [Eq. (\ref{eq:JMAKclassicAvrami})] and non-isothermal [Eq. (\ref{eq:JMAKextendedAvrami})] JMAK model, for film irradiated at (b) $10^{13}$ and (c) $10^{14}$ \ipc. }
\end{figure}

Figure \ref{fig:fig6}(a) compares the temperature dependence of the effective crystalline volume fraction for the as-grown and irradiated films, obtained using Eq.~(\ref{eq:DOC}), and the diffraction data is shown in Figs \ref{fig:fig4}(i,j) and \ref{fig:fig5}(i,j). First, notice that the effective crystalline fraction for the pristine sample is around 0.92 pre-annealing and falls to around 0.82 by the end of the annealing sequence, whereas an $X$ of around 1 would ideally be expected for a perfectly crystalline film. We attribute this reduced extracted crystallinity to artefacts --- the broadening of the diffraction spots and the presence of Kikuchi lines, highlighted in Supplementary Materials Fig. S1. Likely owing to slight movements of the sample, the Kikuchi lines shift, causing an inconsistent broadening of the diffraction spots at different temperatures. Extensive diffraction data collected on the pristine sample is included in the Supplementary Materials.

Before the recrystallization process starts, we see that the sample irradiated at $10^{14}$ \ipc \ exhibits a crystallized volume fraction of approximately 0.22, while the film irradiated with the lower fluence has a significantly higher initial crystallized volume fraction of approximately 0.6. This disparity is consistent with our more qualitative observations from microscopy and diffraction studies, all showing that the higher ion dose leads to more amorphous regions, resulting in a lower crystallized volume fraction. We also see from Fig. \ref{fig:fig6}(a) that after annealing at a maximum temperature of 250$^{\circ}$C, $X$ is nearly equivalent for the film irradiated at $10^{13}$ \ipc \ and the as-grown film. However, the crystallized volume fraction for the sample irradiated at $10^{14}$ \ipc \ reaches a lower value compared to the lower-dosed sample. This suggests that either longer annealing times are required or there are defects that have larger activation energies which require higher temperatures to eliminate.

Figure \ref{fig:fig6}(a) also reveals a distinct difference in the temperature at which $X(T)$ upturns in each sample, showing that recrystallization in the sample irradiated at $10^{14}$ \ipc \ occurs at a higher temperature compared to the one irradiated at $10^{13}$ \ipc.  This shift suggests that the higher density of the amorphous phase leads to greater phase stability and thus requires a higher activation energy for recrystallization. To validate this explanation, further experiments are needed, for example, differential scanning calorimetry to calculate the activation energy\cite{DSCMethods1} using the Kissinger method.\cite{DSCMethod2, DSCMethod3}

Next, from our crystallized volume fraction data, we can apply the JMAK model to determine the Avrami exponent and then extract information regarding the nucleation rate and mechanism from these values.  As shown in Fig.~\ref{fig:fig3}, our heating profile is stepwise, therefore, neither completely isothermal nor non-isothermal. Consequently, we apply both the classic and extended JMAK models, Eq.~(\ref{eq:JMAKclassicAvrami}) and (\ref{eq:JMAKextendedAvrami}), respectively. For the extended model, which considers non-isothermal conditions, we approximate the temperature sweep as linear, for $\beta = 2^{\circ}\text{C/min}$, and use an activation energy of $E_a = 0.1011 \times 10^{-19}$ J/atom for FeGe, based on Ref.~[\onlinecite{FeGeActivationE}].

Figures \ref{fig:fig6}(b, c) show the calculated local Avrami exponents $n(X)$ plotted against the crystallized volume fraction $X$, for the film irradiated at $10^{13}$ \ipc \ and $10^{14}$ \ipc, respectively.  To now resolve the nucleation rate and growth mechanism, we consider that the local Avrami exponent $n(X)$ can be described using the equation $n = a + bc$.\cite{AvramiConstant, NonIsoApplic}  Here, $a \geq 0$ provides information on the relative nucleation rate, where $a < 1$, $a = 0$, and $a > 1$ for a decreasing, constant, or increasing nucleation rate with time, respectively. The term $b$ relates to the growth mechanism, for which $b = 1$ indicates interface-controlled growth and $b = 0.5$ signifies diffuse-controlled growth. Lastly, the term $c$ indicates the growth dimension, therefore can only be 1, 2, or 3.

Based on our data, the possible indices $a$, $b$, and $c$ are not unique. We discuss the possibilities here, which are also summarized in Table \ref{tab:Avrami}. It has been previously reported that crystallization of Fe-Ge films is diffusion controlled,\cite{DiffusionBased} therefore we assume $b = 0.5$. Consequently, under this assumption, the lower limit for $n(X)$ is 0.5, calculated from minimum values for $c=1$ and $a=0$. Therefore, for the film irradiated at 10$^{13}$ \ipc, our $n(X)$ data that can be applied to the model lies within the range $0.64 < X < 0.85$, highlighted in grey shading in Fig. \ref{fig:fig6}(b). According to the classic JMAK model, the $n$ value initially increases to approximately 1.2 before rapidly decreasing to below 1 with increasing temperature and $X$. Similarly, $n(X)$ calculated from the extended JMAK model exhibits comparable behavior, increasing initially, though it peaks at a lower value of 1.0 before decreasing. We then use the average Avrami exponent $\overline{n}$ --- 0.9 and 0.8 for the JMAK and extended JMAK models, respectively --- to determine the indices. For an increasing nucleation rate ($a > 1$) to exist, $n(X)$ must be greater than 1.5 even for the lowest possible $c = 1$, again considering $b=0.5$. Under the constraint of $\overline{n} < 1.0$ and the assumption of $b = 0.5$, $c$ can only be 1 and $a < 1$. Therefore, both models conclude that grain growth in the film irradiated at 10$^{13}$ \ipc \ is diffusion-controlled, one-dimensional growth with a decreasing nucleation rate.

We now consider the film irradiated at 10$^{14}$ \ipc, examining our $n(X)$ data within the range of $0.27 < X < 0.65$, which is highlighted in grey shading in Fig. \ref{fig:fig6}(c). This applicable range was determined using a lower limit of $n(X) = 0.5$. We see that the recrystallization behavior follows a similar pattern to that in the lower-dose film: the nucleation rate decreases since $n(X)$ mostly remains below 1.5. However, the peak $n$ values are higher, reaching 1.6 and 1.34 for the extended and classic JMAK models, respectively. In this case, the $\overline{n}$ for JMAK is 1.1, and for extended JMAK is 1.0. Accordingly, $1.0 < \overline{n} <  1.1$, $a < 1$, and $b = 0.5$ indicate that the grain growth is dominated by diffusion-controlled one-to-two dimensional growth with a decreasing nucleation rate.  

{\renewcommand{\arraystretch}{1.4}
\begin{table}[h]
\begin{ruledtabular}
\begin{tabular}{lccccccc}
        \hspace{-0.0cm}\thead{Irradiation\\\hspace{-0.3cm}Fluence} & \thead{Model} & $X$ & $\overline{n}$ & $a$ & $b$ & $c$ & \thead{Recrystallization \\Mechanism} \\ \hline
$10^{13}$ & Iso & [0.64, 0.85] & 0.9 & [0, 1]  & 0.5 & 1 & 1D, I$\downarrow$ \\
$10^{13}$ & Non-Iso & [0.64, 0.85] & 0.8 & [0, 1]  & 0.5 & 1 & 1D, I$\downarrow$ \\
$10^{14}$ & Iso & [0.28, 0.65] & 1.1 & [0, 1]  & 0.5 & 1, 2 & 1-2D, I$\downarrow$ \\
$10^{14}$ & Non-Iso & [0.28, 0.65] & 1.0 & [0, 1]  & 0.5 & 1, 2 & 1-2D, I$\downarrow$ \\
\end{tabular}
\caption{\label{tab:Avrami} Summary of recrystallization mechanisms showing the irradiation fluence in \ipc, JMAK model applied [isothermal (Iso) or non-isothermal (Non-iso)], range (inclusive, indicated as [min,max]) of extracted crystalline volume fractions $X$, average Avrami exponent $\overline{n}$ within the $X$ data range, possible indices of nucleation (a), growth (b = 0.5 for diffusion controlled), and dimensionalities (c) based on $\overline{n}$, $a$, and $b$. The last column details the recrystallization mechanism associated with these indices, where I$\uparrow$ indicates increasing, I$\downarrow$ indicates decreasing nucleation rate, and the growth dimension is noted as 1D, 2D, or 3D. }
\end{ruledtabular}
\end{table}}

The difference between $n(X)$ for the classic and extended JMAK models reaches a maximum of approximately 16\% in both samples, with all terms remaining within a range that yields the same conclusions regarding nucleation. Both models indicate our irradiated FeGe film experienced a decrease in nucleation rate at all times. This is consistent with the non-zero crystallized volume fraction at the start of annealing because these unaffected crystalline regions can act as nuclei. Notably, the higher dimensional growth observed in the film irradiated at 10$^{14}$ \ipc \ compared to the one irradiated at 10$^{13}$ \ipc \ aligns with the experimental conditions: the higher density of amorphous regions in the former leads to less restricted growth directions.

\section{Conclusions}\label{sec:conclusions}

 We developed a process to tune disorder in epitaxial B20-phase FeGe films, which may be useful for controlling magnetic phenomena in this and other B20-phase materials. For example, irradiating the films creates amorphized regions, and the resulting crystalline-amorphous composite may host skyrmions and antiskyrmions. Annealing progressively recrystallizes the ion-beam-modified films. By monitoring the heating-induced changes in the crystalline structure through in-situ SAED, we determined that as the temperature increases, the crystalline volume fraction is first constant, then sharply increases, before plateauing. We also found that the temperature at which the crystalline volume fraction sharply increases depends on initial structural conditions and that the mostly amorphized FeGe from high fluence irradiation does not fully recrystallize at temperatures up to 250$^{\circ}$C within the duration of our annealing process (104 minutes total, 3 minutes at 250$^{\circ}$C). The primary phase in both recrystallized samples remains as the B20-phase. Lastly, by applying both JMAK models, we determined that the nucleation rate in both samples decreases during recrystallization, which suggests the formation of larger grains with fewer grain boundaries.  Further studies are warranted --- namely, LTEM, XMCD, or neutron scattering --- to definitively identify the magnetic structures that emerge in irradiated FeGe and annealed FeGe. Our results can subsequently contribute to establishing structure-property relationships in the FeGe system for spintronic applications.

\section{Methods}\label{sec:methods}

\subsection{Film Growth}

Our FeGe films on Si (111) substrates were prepared at the Platform for the Accelerated Realization, Analysis, and Discovery of Interface Materials (PARADIM) at Cornell University, applying a process similar to that detailed in Ref. ~\onlinecite{PhysRevMaterials.2.074404} for growth of Mn$_x$Fe$_{1-x}$Ge films. Using molecular beam epitaxy in a Veeco GEN10 system ($2 \times 10^{-9}$ torr base pressure), a FeSi seed layer was first formed by depositing a monolayer of Fe onto a $7 \times 7$ reconstructed Si (111) surface then flash annealing at \ang{500}C. Fe and Ge sources were then co-deposited from 40 cc effusion cells at a rate of 0.5 \AA/s and substrate temperature of \ang{200}C. This resulted in B20-phase, epitaxial FeGe films that were fairly uniform over 1.5 inch substrates.

\subsection{Ion Beam Modification}

The FeGe films were irradiated in a 6 MV HVE EN Tandem Van de Graaff Accelerator at the Ion Beam Lab at Sandia National Laboratories. Prior to the irradiation process, the beam energy and fluences were chosen based on estimations of resulting atomic displacements from Stopping Range of Ion in Matter (SRIM) simulations.\cite{Ziegler2010} The SRIM simulation considered the densities,  displacement energies, lattice energies, and surface energies indicated in Table \ref{tab:SRIM}. FeGe films cleaved into \qtyproduct{6 x 6}{mm} fragments were adhered onto a Si backing plate with double sided carbon tape and mounted into the tandem end station which was pumped to a base pressure of at least \qtyproduct{1e-6}{torr}. Ion irradiation was performed using 2.8 MeV $\mathrm{Au^{4+}}$ ions with an ion beam current of \qtyproduct{100}{nA}, equivalent to an ion flux of \qtyproduct{1.56e11}{ions/cm^2}.

\begin{table}[h]
\begin{ruledtabular}
\begin{tabular}{lrrr}
                  & Fe & Ge & Si \\ \hline
Displacement Energy (eV) & 25 & 15 & 15  \\
Lattice Energy (eV)& 3 & 2 & 2 \\
Surface Energy (eV) & 4.34 & 3.88 & 4.7  \\
Density (\unit{g/cm^3}) & \multicolumn{2}{c}{FeGe:  8.14}  &   2.312 \\
\end{tabular} \caption{\label{tab:SRIM} SRIM simulation parameters. Displacement energies, lattice binding energies, surface binding energies, and densities for Fe, Ge, and Si used in the simulations, based on parameters from Ref. ~(\onlinecite{MaterialsProject_FeGe}).}
\end{ruledtabular}
\end{table}

\subsection{Scanning Transmission Electron Microscopy and Focus Ion Beam Milling}

Lamellae were prepared using a Thermo Fisher Helios G496 UX-focused ion beam (FIB) at PARADIM. Plan-view lamellae were used for SAED and cross-sectional lamellae were fabricated for STEM and EELS measurements. Initially, a protective overlayer
stack of 20 - 30 $\mathrm{nm}$ of carbon and 0.8–1 $\mu$m of Pt was deposited on the sample surface. Subsequently, a Ga-ion beam was used to carve out the thinned lamellae, which were then attached to a needle using a sputtered Pt paste to facilitate transfer onto a TEM grid. Both the cross-section and plan-view lamellae went through successive milling on both sides down to a thickness of 200 - 500 $\mathrm{nm}$ using a 30 keV Ga-ion beam. Lastly, the lamellae were milled at 5 keV until the Pt layer thickness fell below $20 \mathrm{nm}$. Throughout the plan-view FIB thinning process, the film surface and substrate underwent milling at a 2-3$^{\circ}$ angle with respect to the ab-plane, ensuring a continuous gradient across different regions. The STEM images depicted in Fig. \ref{fig:fig1}(c) and Fig. \ref{fig:fig2}(d) as well as the EELS data in Fig. \ref{fig:fig3}(c) were captured using a Thermo-Fisher Scientific Spectra 300 STEM in the PARADIM facility.  

\subsection{Selective Area Electron Diffraction}

The SAED data captured during the in-situ annealing process were collected using a 300 kV Titan Transmission Electron Microscope at Sandia National Laboratories (SNL). The sample was sandwiched onto a single-tilt Gatan heating holder using a screw. Then, after the chamber was pumped down to $1.5\times10^{-9}$ torr, the sample was heated from 50$^{\circ}$C to 250$^{\circ}$C using the stepwise profile depicted in Fig. \ref{fig:fig3}. This heating process alternated between a temperature ramp of $5^{\circ}\mathrm{C/min}$ for 2 minutes and then a 3-minute hold. At each holding step, the temperature was allowed to stabilize for one minute, after which the following data was collected over a period of two minutes: a bright-field (BF) TEM image of the entire sample at a magnification of 3800X, a diffraction pattern using the selected area aperture with sample distance of 480 mm with an area of 0.35 $\mu m^{2}$, and a BF TEM image of where the SAED data was taken, both at 8100X. The exposure time for all the images was 1 s and, throughout the imaging process, the SAED location remained fixed and microscope voltage was maintained at 300 kV.

\section{Supplementary Materials}\label{sec:SM}

See Supplementary Materials for selected area electron diffraction results on the as-grown FeGe film as well as further details regarding our SAED data analysis process.

\section*{Author Declarations}

\subsection*{Conflict of Interest}
The authors have no conflicts to disclose.

\subsection*{Authors' contributions}
S.E. conceived and designed the experiment, with input from D.M., K.H., and T.L. M.B.V. conducted the SRIM simulations and coordinated sample preparation, irradiation, and microscopy based on Hall effect studies.
H. P. grew the FeGe films.
X.Z. and H.Y. prepared the FIB samples. X.Z. collected and analyzed EELS and STEM images with advice from D.M. H.Y. also collected TEM images. 
R.S. and K.H. collected the TEM and diffraction data under various temperatures. J.L. analyzed, interpreted, and modeled the diffraction data. 
S.E. and J.L. wrote the manuscript. 
S.E. and T.L. assisted with general data analysis and interpretation.
All authors commented on the manuscript.

\begin{acknowledgments}
This material is based upon work supported by the National Science Foundation under grants DMR-1905909 and DMR-2330562 at the Colorado School of Mines and University of Washington. 
This work also made use of synthesis and microscopy facilities at the Platform for the Accelerated Realization, Analysis, and Discovery of Interface Materials (PARADIM), which are supported by the National Science Foundation under Cooperative Agreement No. DMR-2039380.
Part of this work was performed at the Center for Integrated Nanotechnologies, a DOE Office of Science User Facility, and Sandia National Laboratories, managed and operated by NTESS, LLC, a wholly owned subsidiary of Honeywell International, Inc., for the U.S. DOE’s National Nuclear Security Administration. The views expressed in the article do not necessarily represent the views of the U.S. DOE or the United States Government.
\end{acknowledgments}

\section*{Data Availability Statement}

The data that support the findings of this study are available on Mendeley Data (doi.org/10.17632/xd8s497nsz.3) as a zip file. This includes .csv files containing the data used in figures, microscopy images (.tif files), python code used for the analysis, and an Origin file (.opju) containing all figures and data spreadsheets, which can be opened using Origin Viewer, a free application that permits viewing and copying of data contained in Origin project files.

\bibliography{aipsamp}

\end{document}


\preprint{AIP/123-QED}

\title{Supplemental Materials: Structural Properties and Recrystallization Effects in Ion Beam Modified B20-type FeGe Films}
%
%

\author{Jiangteng Liu}
\affiliation{Department of Electrical \& Computer Engineering, University of Washington, Seattle, WA 98195}%

\author{Ryan Schoell}%
\affiliation{Sandia National Laboratories, Albuquerque, NM 87123}%

\author{Xiyue Zhang}%
\affiliation{Department of Applied Physics, Cornell University, Ithaca, NY 14853}%

\author{Hongbin Yang}%
\affiliation{Department of Applied Physics, Cornell University, Ithaca, NY 14853}%

\author{M. B. Venuti}
\affiliation{Department of Physics, Colorado School of Mines, Golden, CO 80401}%
 
\author{Hanjong Paik}
\affiliation{Platform for the Accelerated Realization, Analysis, and Discovery of Interface Materials, Cornell University, Ithaca, NY 14853}
\affiliation{School of Electrical \& Computer Engineering, University of Oklahoma, Norman, OK 73019}
\affiliation{Center for Quantum Research and Technology, University of Oklahoma, Norman, OK 73019}

\author{David Muller}%
\affiliation{Department of Applied Physics, Cornell University, Ithaca, NY 14853}%

\author{Tzu-Ming Lu}%
\affiliation{Center for Integrated Nanotechnologies, Sandia National Laboratories, Albuquerque, NM 87123}%

\author{Khalid Hattar}%
\affiliation{Center for Integrated Nanotechnologies, Sandia National Laboratories, Albuquerque, NM}%
\affiliation{Department of Nuclear Engineering, University of Tennessee, Knoxville, TN 37996} 

\author{Serena Eley}
\affiliation{Department of Electrical \& Computer Engineering, University of Washington, Seattle, WA 98195}
\affiliation{Department of Physics, Colorado School of Mines, Golden, CO 80401}
\email[Corresponding author: ]
{serename@uw.edu}

\date{\today}

\maketitle

\setcounter{figure}{0}
\makeatletter 
\renewcommand{\thefigure}{S\@arabic\c@figure}
\makeatother

\subsection{Selected Area Electron Diffraction (SAED) Results on Pristine FeGe Film}

\begin{figure*}[h]
\centering
\includegraphics[width=0.7\linewidth]{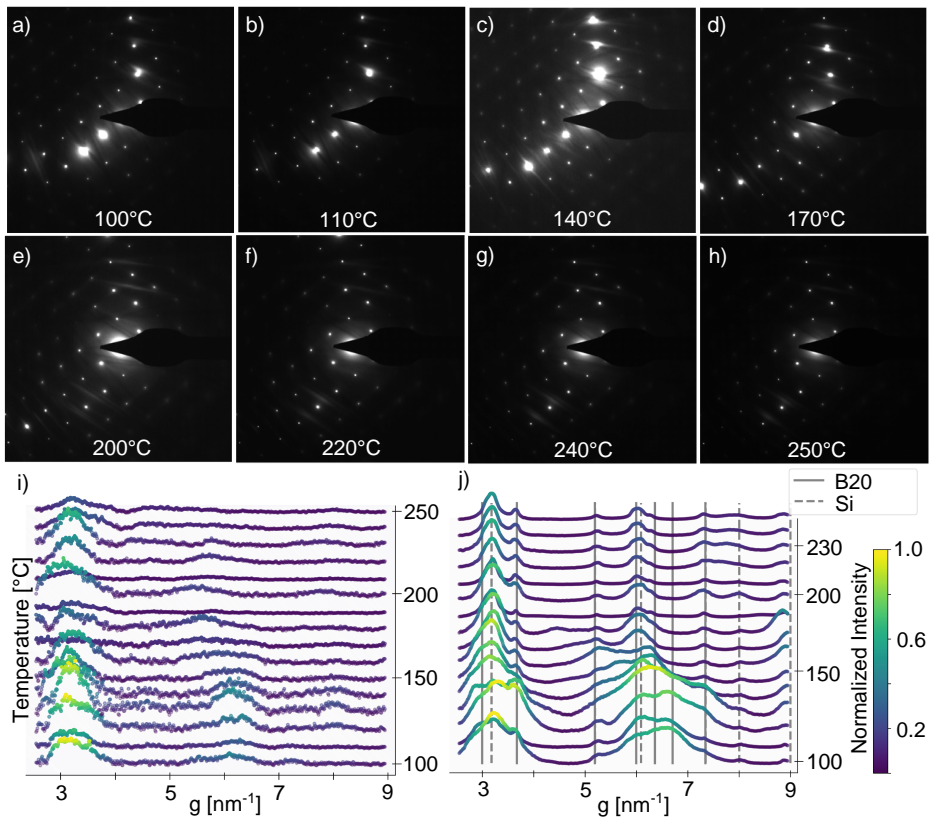}
\caption{\label{fig:figS1} Selected area electron diffraction images of pristine FeGe film. (a-h) Raw electron diffraction images at different stages of annealing, at the indicated temperatures. Extracted intensity (color scale) versus reciprocal space parameter $g$ for the (i) amorphous regions and (j) crystalline regions at different annealing temperatures. Curves offset for clarity.  The solid grey lines indicate the expected position of peaks associated with the B20-phase phase.}
\end{figure*}

 In-situ recrystallization was conducted on a pristine FeGe lamella to serve as a control. The goal was to  understand the effects of annealing sample that had not undergone ion beam modification and to test our analysis procedure determining the effective crystalline volume fraction. Comparison with the irradiated samples helps us distinguish the effects of ion-beam modification from the inherent material properties. Figure \ref{fig:figS1}(a-h) presents the raw SAED images of the pristine FeGe lamella, with each panel showing results at a fixed location while heating at the indicated temperature. As expected for a pristine crystalline sample, the SAED images show no visible diffuse ring, confirming that the sample is purely crystalline. 
 
 Kikuchi lines, which appear as pairs of parallel dark and bright lines, are present in the diffraction images and are particularly visible in Fig \ref{fig:figS1}(c). These patterns are typically observed in thick, well-crystallized samples and occur due to electrons scattered inelastically at small angles with minimal energy loss.\cite{Kikuchiline} In addition to the Kikuchi lines, inconsistent broadening of the diffraction spots is also observed. This broadening is likely caused by slight movements of the sample during the experiment, leading to the shifting of Kikuchi lines. When a Kikuchi line aligns with a diffraction spot, broadening occurs, affecting the precision of the diffraction data. Figures~\ref{fig:figS1}(i) and \ref{fig:figS1}(j) display the resulting separated 1D profiles for the amorphous and crystalline components, respectively. The intensity of the amorphous component remains unchanged during annealing, verifying that the experiment did not induce any recrystallization. For the crystalline component shown in Fig. \ref{fig:figS1}(j), only the Si and B20 FeGe phases have been identified. Additionally, a broadening of the diffraction peaks due to the Kikuchi lines is observed, but no significant changes in intensity are detected. This further confirms that the pristine sample is fully crystalline and has not undergone any phase transformation during the experiment. However, this broadening results in a decrease in the extracted crystalline volume fraction $X$, applying our analysis procedure. This allows us to set a lower bound for what value of $X$ may actually be indicative of full crystallization, reduced due to artefacts such as Kikulchi lines, in the annealed, ion-beam modified sample.

\subsection{Separation of SAED Data into Crystalline and Amorphous Components}

\begin{figure*}[h]
\centering
\includegraphics[width=0.65\linewidth]{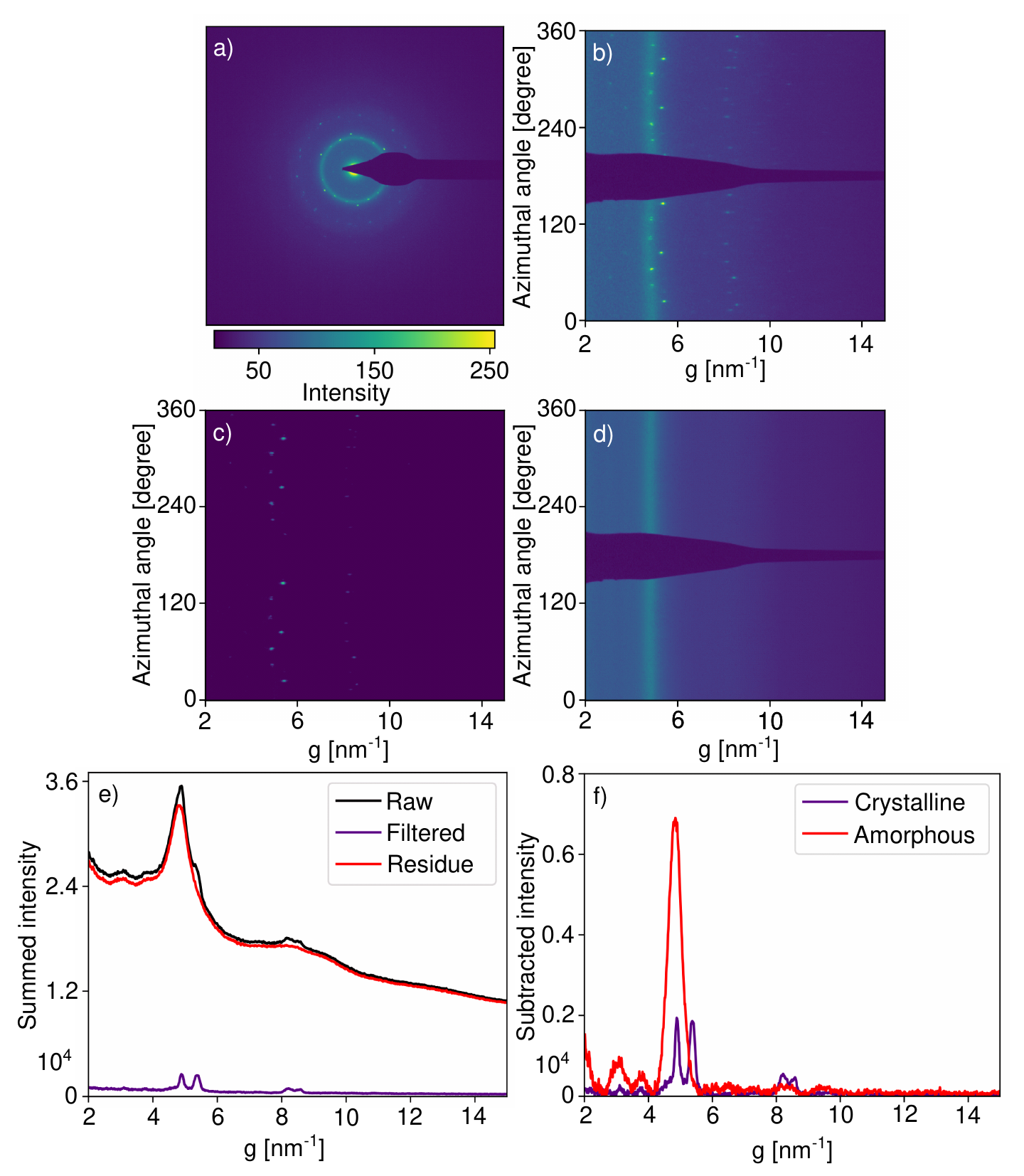}
\caption{\label{fig:figS2} The different stages during the signal separation of crystalline and amorphous components from raw SAED images. The example data were collected on the film irradiated at $10^{14}$ ions/cm² while annealing at 100$^{\circ}$C. (a) The original 2D diffraction image in .tif format obtained from the Titan Transmission Electron Microscope. (b) 2D graph of the raw image after using the 2D integration/caking function in the pyFAI package, with rows representing the azimuthal angle (y-axis) and columns representing the reciprocal space parameter $g = 1/d$ (x-axis). Using the median threshold filter, the data is separated into curves containing a baseline and the (c) crystalline component, and (d) amorphous components. The baseline originates from thermal diffuse scattering\cite{Thermaldiffuse}, (e) 1D diffraction plot obtained by summing the azimuthal angles for the raw, filtered, and residue data (raw minus filtered). (f) The filtered and the residue data after applying the Asymmetric Least Squares baseline subtraction estimation, which now becomes the crystalline and amorphous components.}
\end{figure*}

Raw diffraction data from the Titan Transmission Electron Microscope is formatted as .tif images. To separate the diffraction traces associated with the crystalline versus amorphous regions in the sample, we followed the process described here. The example data are for the sample irradiated with a fluence of $10^{14}$ \ipc \ after annealing at 100$^{\circ}$C, shown in Fig. \ref{fig:figS2}(a).

First, four standard Au diffraction images were taken to calibrate the instrument. Second, the scikit-ued package \cite{RenedeCotret2016, RenedeCotret2018} was used to locate the beam center and choose the number of bins to use for the reciprocal space parameter $g$. Third, 2D integration was performed on each image using the pyFAI packages, \cite{PyFAI} which converts the .tif images into a two-dimensional array in which the rows are the azimuthal angle and the columns are the reciprocal space parameter $g$ shown in Fig \ref{fig:figS2}(b). For the amorphous phase, we assumed that the background scattering at each reciprocal space parameter $g$ was uniform across all azimuthal angles. Therefore, the signal separation was performed on each column. Here, any intensity value that is less than the third quartile is considered as background and amorphous phases thus were filtered. Furthermore, to avoid any residual background signal remaining in the filtered data, the median of the background signal was also calculated and removed from the filtered data. This process results in two 2-dimensional arrays in which the filtered data contains the crystalline components Fig \ref{fig:figS2}(c), and the residue contains the amorphous component Fig \ref{fig:figS2}(d). Fourth, all azimuthal angles are summed to produce 1D diffraction data shown in Fig \ref{fig:figS2}(e). Last, we apply the Asymmetric Least Squares baseline estimation function in Python to remove the background.\cite{ALSMethod} This is shown in Fig \ref{fig:figS2}(f), which is the processed final data used to create figures in the main text.

\subsection{Phase Identification}

\begin{figure*}[h]
\centering
\includegraphics[width=0.9\linewidth]{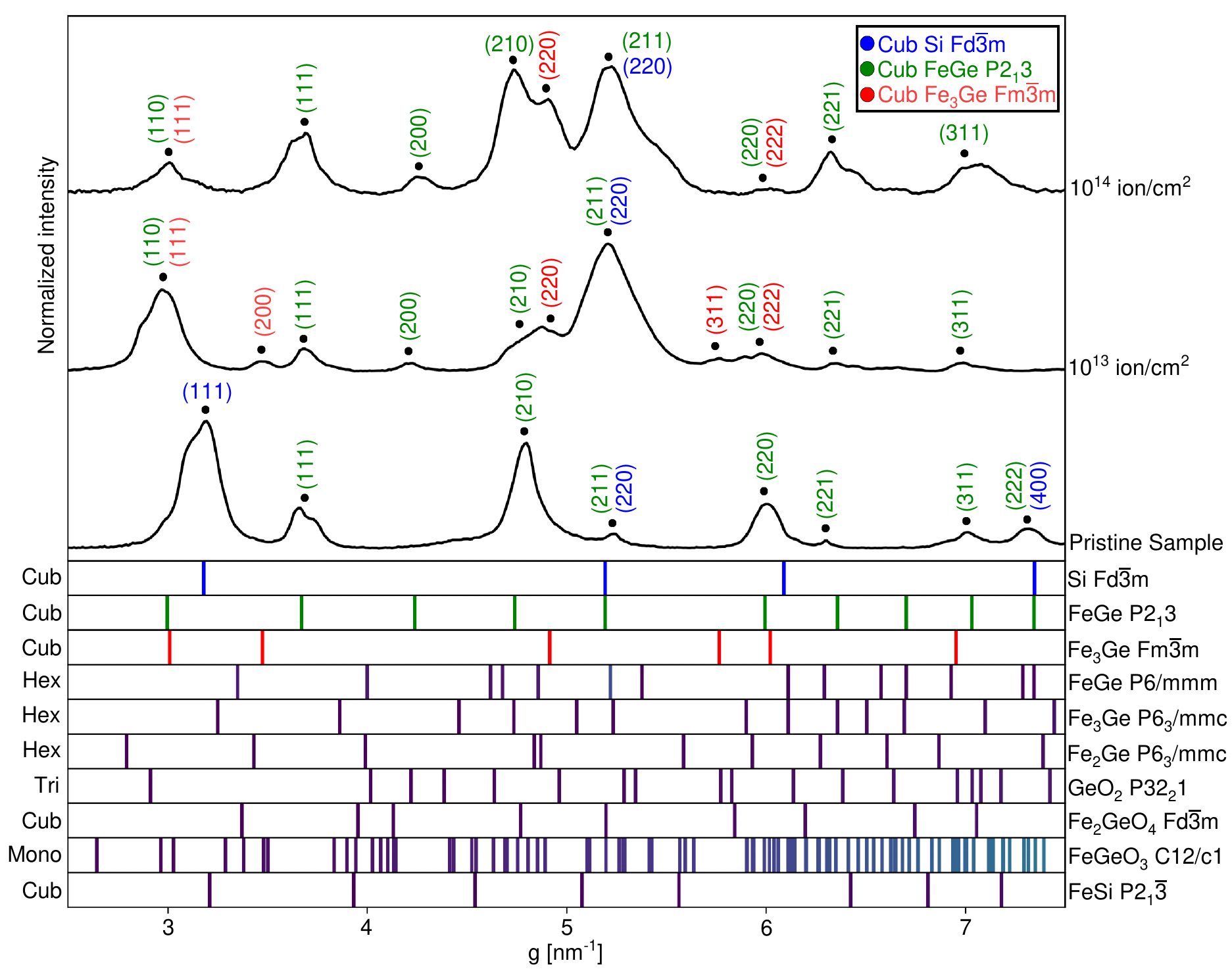}
\caption{\label{fig:figS3} Example phase identification for annealed FeGe films. The top curve shows the data for the pristine FeGe film after annealing at 250 $^{\circ}$C; the middle curve is for the FeGe film irradiated at $10^{13}$ \ipc \ after annealing at 220 $^{\circ}$C; and the bottom curve is for the FeGe film irradiated at $10^{14}$ \ipc \ after annealing at 250 $^{\circ}$C. Diffraction peaks are indexed to their corresponding crystallographic planes. Identifiable phases include cubic Si (blue), cubic FeGe (green), and cubic Fe$_{3}$Ge (red); their calculated diffraction pattern peak positions are shown below using the same color coding. Some common cubic (Cub), hexagonal (Hex), triclinic (Tri), and monoclinic (Mono) phases from Fe-Ge and Fe-Ge-O systems are also included as references.}
\end{figure*}

An extensive phase identification process is performed on the SAED patterns using the Material Database.\cite{MaterialsProject_FeGe} We focused on all the Fe-Ge and Fe-Ge-O systems as well as some other possible phases, e.g. GeO$_{3}$, FeSi. Figure \ref{fig:figS3} presents a detailed, indexed 1D diffraction plot example for all three annealed samples.

The search result for the pristine sample reveals that only cubic Si and cubic FeGe (B20) phases are present. Regarding the annealed sample irradiated at $10^{13}$ \ipc, cubic Si, cubic FeGe (B20), and cubic Fe$_{3}$Ge phase were identified. Lastly, diffraction data for the annealed sample irradiated at $10^{14}$ \ipc shows evidence of cubic Si and cubic FeGe (B20) phases, as well as traces of cubic Fe$_{3}$Ge, e.g. (220) peaks. Unlike the film irradiated at the lower fluence, there are no other peaks, for example (200) and (311), to further confirm that the cubic Fe$_{3}$Ge phase is present. Fig. \ref{fig:figS3} also presents examples of calculated phase patterns for other common Fe-Ge and Fe-Ge-O systems that we considered, but did not match our data.

\clearpage
\bibliography{smbib}